\documentclass[prb,showpacs]{revtex4}
\usepackage{graphicx}
\usepackage{subfigure}
\usepackage{amssymb}
\usepackage{amsmath}
\usepackage{bm}
\usepackage{latexsym}
\usepackage{colordvi,epsfig,color,amsmath,amssymb}

\begin{document}

\begin{center}
\baselineskip 40pt

\vskip 2cm


{\Large {\bf Does temperature favor quantum coherence of a dissipative two-level system?}}



\vskip 1cm

{\large Zhiguo L\"{u} and  Hang Zheng}

{Department of Physics, Shanghai Jiao Tong University, Shanghai
200240, China}

{\bf Abstract}
\end{center}

\baselineskip 20pt

The quantum dynamics of a two-level system coupled to an Ohmic spin-
bath is studied by means of the perturbation approach based on a
unitary transformation. A scattering function $\xi_k$ is introduced
in the transformation to take into account quantum fluctuations. By
the master equation within the Born approximation, nonequilibrium
dynamics quantities are calculated. The method works well for the
coupling constant $0 < \alpha < \alpha_c$ and a finite bare
tunneling $\Delta$. It is found that (i) only at zero temperature
with small coupling or moderate one does the spin-spin-bath model
display identical behavior as the well known spin-boson-bath model;
(ii) in comparison with the known results of spin-boson-bath model,
the coherence-incoherence transition point, which occurs at
$\alpha_c={\frac{1}{2}}[1+\eta\Delta/\omega_c]$, is temperature
independent; (iii) the nonequilibrium correlation function
$P(t)=\langle\tau_z(t)\rangle$, evolves without temperature
dependence while $\langle\tau_x(t)\rangle$ depends on temperature.
Both $P(t)$ and $\langle\tau_x(t)\rangle$ not only satisfy their
initial conditions, respectively, and also have correct long time
limits. Besides, the Shiba's relation and sum rule are exactly
satisfied in the coherent regime for this method. 
Our results show that increasing temperature does not help the system suppress
decoherence in the coherent regime, i.e., finite temperature does not favor the coherent dynamics in
this regime. Thus, the finite-temperature dynamics induced by two kinds of baths spin-bath and
boson-bath exhibit distinctly different physics.


\vskip 1cm


\pagebreak

\baselineskip 20pt

\section{Introduction}

The dynamics of a dissipative two-level system has attracted
extensive studies in last decades\cite{rmp,book}, since it can be
used to describe a large number of different physical and chemical
processes. The interaction between the system and its environment
gives rise to decoherence and dissipation which are also the major
stumbling block to quantum computation and quantum communication.
Generally, there are two kinds of quantum environments, one is
boson-bath modeled by a set of oscillators (delocalized modes such
as phonons and photons), the other is two-level systems (TLSs)
reservoir or spin-bath (localized modes such as defects, impurities,
nuclear and paramagnetic spins)\cite{stamp}. Usually, people use the
spin-boson-bath (SBB) model
\begin{eqnarray}
&&H_{SBB}=-{\frac{1}{2}}\Delta \tau_{x} +\sum_{l}\omega
_{l}b^{\dag}_lb_l+{\frac{1}{2}}\sum_{l}g_{l}(b^{\dag}_l+b_l)\tau_z
,
\end{eqnarray}
to take into account the system-bath interaction[4-26], where a
bosonic heat bath consisting of an infinite number harmonic
oscillators (denoted by $b^{\dag}_l$'s and $b_l$'s) constitutes the
environment of a quantum TLS (denoted by the Pauli matrices $\tau_x$
and $\tau_z$). Experimentally, the spin-bath plays an important role
on the decoherence in magnetic cluster and semiconductor qubits at
low temperature with some interesting features\cite{SiGaAs}. For
example, GaAs quantum dot electron spin qubit, a candidate of solid
state quantum computation, loses its quantum memory due to its
coupling with the surrounding nuclear spin environment being the
spin-bath. The abundance of spin-baths in real systems urgently
necessities an understanding of effects on decoherence. Another
simple model for spin-bath is proposed, which is the so-called
spin-bath composed of an infinite number of TLSs without mutual
interaction\cite{nitzan,cald,Shao,Makri,Golosov3,Gelman}. The
Hamiltonian of a TLS coupled with a dissipative spin-bath
(spin-spin-bath, SSB) reads
\begin{eqnarray}
&&H=-{\frac{1}{2}}\Delta \tau_{x} -{1\over 2}\sum_{l}\omega
_{l}\sigma^l_z -{\frac{1}{2}}\sum_{l}g_{l}\sigma^l_x\tau_z.
\end{eqnarray}
Here $\tau_x$ and $\tau_z$ are Pauli matrices to describe the TLS,
$\sigma^l_x$ and $\sigma^l_z$ are Pauli matrices for the $l$-mode of
spin-bath\cite{cald}. $\Delta$ is the bare tunneling matrix,
$\omega_l$ the frequency for the $l$-mode of the bath, and $g_{l}$
the coupling constant. The coupling between the TLS and its
environment is characterized by a spectral density
$J(\omega)=\sum_{l}g^2_{l}\delta(\omega-\omega_l)=2\alpha\omega
\theta(\omega_c-\omega)$ with the dimensionless coupling strength
$\alpha$, the upper cutoff $\omega_{c}$ and the step function
$\theta(x)$.

Most researchers are interested in an open system with a small
number of degrees of freedom (such as a two-level system or an
oscillator) in contact with a ¡®¡®bath¡¯¡¯ of a complex nature,
whose number of degrees of freedom tends to infinity. The evolution
and its properties of the open system are determined by the coupling
to the bath. Do the boson-bath and spin-bath have the same effects
on the dynamics of TLS?  What is the difference between the coherent
dynamics of SBB and SSB? At first glance, differences of decoherence
to the open system could be attributed to intrinsic energy level
structure of heat baths.  The boson-bath may be treated as an
infinitely large reservoir of energy, since the number of
oscillators is infinite and every harmonic oscillator can be excited
to the equally distributed states without upper limit. On the other
hand, the spin-bath may also be treated as an reservoir because the
number of spins in the bath is infinite, but there is only a single
excited state for every bath spin. Thus, the distinct physics
between them ascribe to their underlying structures of bath and
continue to attract much attention from both theoretical and
experimental sides\cite{breuer,hamdouni,smith,schloss}.

The SSB model was studied by approximate analytical and numerical
methods, such as Kubo's cumulant expansion method\cite{nitzan}, the
perturbation theory\cite{cald}, the resolvent operator
approach\cite{Shao}, the numerical path integral method\cite{Makri}.
The main theoretical interest is to understand how the environment,
the spin-bath, influences the dynamics of the TLS and, in
particular, to discuss the common features and the main differences
between the dissipative roles played by the spin-bath and
boson-bath. Some studies show that decoherence is partially
suppressed by increasing the temperature of spin-bath and
temperature plays, though weakly, a positive role in maintaining
coherent dynamics\cite{Shao,Makri}, which is in contrast with the
conclusion of the cumulant expansion method that the decay rate of
TLS is temperature independent\cite{nitzan}.

For instance, the study of noninteracting blip approximation
(NIBA) can give the population difference
$P(t)_{NIBA}$\cite{Makri,Dekker}. Its Laplace transform $P(s)$ for
spin-bath is given
\begin{eqnarray}
&&P(s)=[s+f(s)]^{-1},
\end{eqnarray}
where
\begin{eqnarray}
&&f(s)=\Delta^2\int_0^{\infty}\cos(Q_1(t))\exp(-st-Q_2(t)) dt,
\end{eqnarray}
and the function $Q_1$ and $Q_2$ are given by the relations
\begin{eqnarray}
Q_1 &=& \int_0^{\infty} \sin(\omega t) \tanh(\frac{\omega}{2T})
\frac{J(\omega)}{\omega^2}
d\omega ,\\
Q_2 &=& \int_0^{\infty} [1-\cos(\omega t)]
\frac{J(\omega)}{\omega^2} d\omega.
\end{eqnarray}
As shown in Ref. \cite{Makri}, their simulations as well as the
solution of the NIBA equations, indicated that the diffuse
coherent-incoherent boundary shifts to stronger coupling as the
temperature is raised, which is contrary to the known behaviors for
boson-bath(With increasing temperature, $\alpha_c$ decreases
quickly). It stands a striking contrast against the general belief
that the heat bath with infinite degrees of freedom (thermal
reservoir) leads to the dissipation and decoherence of the open
system and the increase of temperature quenches or does not favor
the coherence. Despite of some different arguments for the
coherent-incoherent transition in this model, these works highlight
that the decoherence of open system depends crucially on the
underlying nature of spin bath. Thus we would reconsider the topic
by an analytical method based on a unitary transformation.

In this work we present an analytical approach for calculating the
dissipative quantum dynamics of the SSB model. It works well for the
coupling constant $0 < \alpha < \alpha_c$ and a finite bare
tunneling $\Delta$, and could reproduce nearly all exact results
obtained by various analytical and numerical methods\cite{zheng,lu}.
It could explain the crossover between the coherent oscillation and
the incoherent behaviors and also allow us to resolve the
controversial claims in the literature. Throughout this paper we set
$\hbar =1$ and $k_{B}=1$.

\section{Unitary transformation}

A unitary transformation, which is defined as $H^{\prime }=\exp
(S)H\exp (-S)$, is applied to $H$ and its aim is to take into
account the correlation between the TLS and its bath. To this end,
the following form for the generator is proposed\cite{zheng,lu},
\begin{equation}\label{S-form}
S=\tau_z\sum_{l}\frac{g_{l}}{2\omega _{l}}\xi _{l}i\sigma^l_{y} .
\end{equation}
Here a $l$-dependent function $\xi _{l}$ is introduced in $S$ and
its form will be determined later.

The transformation can be done to the end and the result is
\begin{eqnarray}\label{H'}
&&H^{\prime }=\exp(S)H\exp(-S) \nonumber\\
&&=-{\frac{1}{2}}\Delta \tau_{x}\cosh[\sum_lv_li\sigma^l_y]
-{\frac{1}{2}}\Delta i\tau_{y}\sinh[\sum_lv_li\sigma^l_y] \nonumber\\
&&-{1\over 2}\sum_{l}[\omega _{l}\cos(v_l)+g_l\sin(v_l)]\sigma^l_z
+{\frac{1}{2}}\tau_z\sum_{l}[\omega_l\sin(v_l)-g_{l}\cos(v_l)]\sigma^l_x
,
\end{eqnarray}
where $v_l=g_l\xi_l/\omega_l$. The transformed Hamiltonian $H'$
may be divided into three parts,
\begin{eqnarray}
&&H^{\prime }=H_{0}^{\prime }+H_{1}^{\prime }+H_{2}^{\prime }, \nonumber\\
&&H_{0}^{\prime }=-{\frac{1}{2}}\eta\Delta \tau_{x}-{1\over
2}\sum_{l}\omega'_{l}\sigma^l_z ,
\end{eqnarray}
with a renormalized frequency for the bath,
\begin{eqnarray}\label{w'}
&&\omega'_l=\omega _{l}\cos(v_l)+g_l\sin(v_l),
\end{eqnarray}
and a renormalized factor for tunneling
\begin{eqnarray}\label{eta}
&&\eta =\mbox{Tr}_B(\rho_B \cosh[\sum_lv_li\sigma^l_y])=
\exp(\sum_{l}\ln[\cos(v_l)]).
\end{eqnarray}
Here $\rho_B=\exp(-\beta H_B)/\mbox{Tr}\exp(-\beta H_B)$ is the
equilibrium density operator of bath spins ($H_B=-{1\over
2}\sum_l\omega'_l\sigma^l_z$) and Tr$_B$ is the trace operation
with respect to the bath. $H^{\prime}_0$ is the unperturbed part
of $H^{\prime}$ and, obviously, it can be solved exactly. The
eigenstate of $H_{0}^{\prime}$ is a direct product: $|s\rangle
|\{\pm 1_l\}\rangle$, where $|s\rangle $ is the eigenstate of
$\tau_x$: $ |s_{1}\rangle ={\frac{1}{\sqrt{2}}}\left(
\begin{array}{c}
1 \\
1
\end{array}
\right) $ or $|s_{2}\rangle ={\frac{1}{\sqrt{2}}}\left(
\begin{array}{c}
1 \\
-1
\end{array}
\right) $, and $|\{\pm 1_l\}\rangle$ is the eigenstate of bath
spins: $\{\pm 1_l\}$ means that the eigenvalue of $\sigma^l_z$ is
$+1$ or $-1$. In particular, $|\{+1\}\rangle$ is the vacuum state
with eigenvalue of every $\sigma^l_z$ being $+1$. Then, the ground
state of $H_{0}^{\prime}$ is $|g_{0}\rangle =|s_{1}\rangle
|\{+1\}\rangle $.

The other terms in $H'$ are
\begin{eqnarray}
&&H_{1}^{\prime
}={\frac{1}{2}}\tau_{z}\sum_{l}\sigma^l_x[\omega_l\sin(v_l)-g_{l}\cos(v_l)]
 -{\frac{1}{2}}\eta \Delta i\tau_y\sum_{l}\sin(v_l)i\sigma^l_y, \\
&&H_{2}^{\prime
}=-{\frac{1}{2}}\Delta\tau_x\left(\cosh[\sum_lv_li\sigma^l_y]
-\eta \right)  \nonumber \\
&&-{\frac{1}{2}}\Delta i\tau_y\left(\sinh[\sum_lv_li\sigma^l_y]
 -\eta \sum_l\sin(v_l)i\sigma^l_y \right),
\end{eqnarray}
$H_{1}^{\prime }$ and $H_{2}^{\prime }$ are treated as
perturbation and they should be as small as possible. For this
purpose $\eta$ is determined in Eq.(\ref{eta}) to make
$\mbox{Tr}_B(\rho_B H'_2)=0$. Besides, $\xi_l$ is determined as
\begin{eqnarray}\label{xi}
&&v_l=\frac{g_l}{\omega_l}\xi_l=\tan^{-1}\frac{g_l}{\omega_l+\eta\Delta
},
\end{eqnarray}
and because of this definition one obtains
\begin{eqnarray} \label{H1'}
&&H_{1}^{\prime }=-{\frac{1}{2}}\sum_{l}\frac{\eta\Delta
g_l}{[(\omega_l+\eta\Delta)^2+g^2_l]^{1/2}}\left(\tau_z\sigma^l_x+i\tau_y
i\sigma^l_y\right) \nonumber\\
&&=-\sum_l\frac{\eta\Delta
g_l}{[(\omega_l+\eta\Delta)^2+g^2_l]^{1/2}}
\left[\tau_+\sigma^l_-+\tau_- \sigma^l_+ \right].
\end{eqnarray}
Here we use the following definition for spin operators:
$\sigma^l_+={1\over 2}(\sigma^l_x-i\sigma^l_y)$, $\sigma^l_-={1\over
2}(\sigma^l_x+i\sigma^l_y)$, $\tau_+={1\over 2}(\tau_z+i\tau_y)$,
$\tau_-={1\over 2}(\tau_z-i\tau_y)$, since the ground state of $l$th
bath spin is $\sigma^l_z=+1$ but its excited state is
$\sigma^l_z=-1$. It is easy to check that $H_{1}^{\prime
}|g_{0}\rangle =0$ because $H'_1$ contains the rotating-wave terms
only. The counter-rotating-wave terms in original Hamiltonian
contribute to the renormalization in $H'_0$ and its form in Eq. (12)
disappears because of the functional form of $\xi_l$ in
Eq.(\ref{xi}). At the same time, the bare coupling $g_l$ in $H$ has
changed to an effective coupling in $H'_1$. These are essential in
our approach.

Substituting (\ref{xi}) into (\ref{w'}) and (\ref{eta}), one obtains
the renormalized frequency
\begin{eqnarray}
&&\omega'_l=\frac{\omega_l(\omega_l+\eta\Delta)+g^2_l}{\sqrt{(\omega_l+\eta\Delta)^2+g^2_l}}
\end{eqnarray}
and the renormalized factor of tunneling
\begin{eqnarray} \label{etach}
&&\eta =\exp\left(-{1\over
2}\sum_{l}\ln\left[1+\frac{g^2_l}{(\omega_l+\eta\Delta)^2}\right]\right).
\end{eqnarray}
In our treatment $H'_0$ is the unperturbed Hamiltonian with the
renormalized parameters. $H'_1$ is the perturbation which contains
flip(transition) of single bath spin. $H'_2$ will be neglected
because it involves multi-flip (multi-transition) of two or more
bath spins and its contribution to physical quantities is $O(g^4_l)$
and higher. Thus, in the following treatment, $H' \approx
H'_0+H'_1$.

The renormalized factor of tunneling $\eta$ is important to
determine the physical property of the coupling system. In order to
make the summation over $l$ of physical quantities, without loss of
generality, a constant density of state is introduced,
$\rho(\omega_l)=\rho_0$:
\begin{eqnarray}
&&\sum_lf(\omega_l,g^2_l)=\int^{\omega_c}_{0}d\omega_l\rho_0 f(\omega_l,g^2_l),\\
&&g^2_l=2\alpha\omega_l/\rho_0,
\end{eqnarray}
where $f(\omega_l,g^2_l)$ is any function of $\omega_l$ and $g^2_l$.
It can be checked that the above treatment is in agreement with the
Ohmic spectral density $J(\omega)$. In following calculations the
dimensionless quantity $\rho_0\omega_c$ is treated as the total
number of bath spins which goes to infinity in the thermodynamic
limit. For example, the summation in Eq.(\ref{etach}) can proceed as
\begin{eqnarray} \label{eta-s}
&&\eta =\exp
\left(-\alpha\int^1_0\frac{xdx}{(x+\eta\Delta/\omega_c)^2}\right)=\exp
\left(-\alpha\ln\frac{\omega_c+\eta\Delta}{\eta\Delta}+\frac{\alpha\omega_c}{\omega_c+\eta\Delta}\right),
\end{eqnarray}
where $x=\omega_l/\omega_c$. In the scaling limit
$\Delta\ll\omega_c$ one can get an explicit solution $\eta=
(e\Delta/\omega_ c)^{\alpha/(1-\alpha)}$, which leads to a
localization point at $\alpha=1$, $\eta=0$ for $\alpha>1$.
As the coupling increases, $\eta$ decreases
smoothly to zero.

Eq.(\ref{eta-s}) is exactly the same as that of the SBB model (Ohmic
bath) for ground state at $T=0$ \cite{zheng},
\begin{eqnarray}
\eta_{B} =\exp\left(-{1\over
2}\sum_{l}\frac{g^2_l}{(\omega_l+\eta\Delta)^2}\coth(\frac{\omega_l}{2T})\right).
\end{eqnarray}
$\eta$ of the SSB model is temperature independent but $\eta_{B}$ is
temperature dependent. That is because every bath spin has only a
single excited state but every oscillator in the boson-bath has
infinite excited states. In addition, it is seen that $\eta_{B}$
decreases with the increase of temperature. In other words, the
raise of temperature of bosonic bath leads to the faster loss of
coherence.

As the transformation in Eq.(\ref{H'}) has been done without
approximation, one can calculate the upper bound of the ground state
energy of the coupling system by
\begin{eqnarray}
&&E_g/\omega_c=-{\frac{1}{2}}\eta\Delta/\omega_c -{1\over
2}\sum_{l}[\omega'_{l}-\omega_l]/\omega_c\nonumber\\
&&=-{\frac{1}{2}}\eta\Delta/\omega_c - {\alpha\over 2}\int^1_0 dx
\frac{x^2+2x\eta\Delta/\omega_c}{(x+\eta\Delta/\omega_c)^2}=-{\frac{1}{2}}\eta\Delta/\omega_c
- {\alpha\over 2}\frac{1}{1+\eta\Delta/\omega_c}.
\end{eqnarray}

Note that the ground state of $H$ is $\exp(-s) |g_{0}\rangle$, it is
the ground state of interacting system. The effect of fluctuating
environment has been taken into account in the treatment.

\section{Non-equilibrium Correlation function}

The density operator in $\rm Schr\ddot{o}dinger$ representation is
$\rho_{SB}(t)$ with Hamiltonian $H$, where the subscript SB
indicates that it is the density operator for the coupled two-level
system and bath. For transformed Hamiltonian $H'$ the density
operator is $\rho'_{SB}(t)=e^S \rho_{SB}(t) e^{-S}$. We treat $H'_0$
as the unperturbed Hamiltonian (Eq.(5)) and the density operator in
the interaction picture is
\begin{eqnarray}
&&\rho^{\prime
I}_{SB}(t)=\exp(iH'_0t)\rho'_{SB}(t)\exp(-iH'_0t).
\end{eqnarray}
The equation of motion for $\rho^{\prime I}_{SB}(t)$ is\cite{scu}
\begin{eqnarray} \label{rho-motion}
&&\frac{d}{dt}\rho^{\prime I}_{SB}(t) =-i[H'_1(t),\rho^{\prime
I}_{SB}(t)].
\end{eqnarray}
$H'_1(t)$ is the perturbation $H'_1$ (Eq.(15)) in the interaction
picture,
\begin{eqnarray}
&&H'_{1}(t)=-\sum_lV_l
\left[e^{-i(\omega_l'-\eta\Delta)t}\tau_+\sigma^l_-+e^{i(\omega_l'-\eta\Delta)t}\tau_-
\sigma^l_+ \right].
\end{eqnarray}
where $V_l=\eta\Delta g_l/[(\omega_l+\eta\Delta)^2+g^2_l]^{1/2}$.
Our procedure for solving the equation is to write\cite{scu}
\begin{eqnarray}
&&\rho^{\prime I}_{SB}(t)=\rho^{\prime I}_{S}(t)\rho_B +\rho^{\prime
I}_c(t),
\end{eqnarray}
where $\rho^{\prime I}_{S}(t)=\mbox{Tr$_B$}\rho^{\prime I}_{SB}(t)$
is the reduced density operator for TLS. Here $\rho^{\prime I}_c(t)$
is the cross term of system and bath operators representing the
correlation because of their interaction, which is the order of
$g_l$ and higher. $\mbox{Tr$_B$} \rho^{\prime I}_c(t)=0$. Now
Eq.(\ref{rho-motion}) can be integrated as
\begin{eqnarray}
&&\rho^{\prime I}_S(t)\rho_B-\rho^{\prime
I}_S(0)\rho_B+\rho^{\prime I}_c(t)-\rho^{\prime I}_c(0)\nonumber\\
&&=-i\int^t_0 dt'[H'_1(t'),\rho^{\prime I}_S(t')\rho_B]-i\int^t_0
dt'[H'_1(t'),\rho^{\prime I}_c(t')].
\end{eqnarray}
After applying the Tr$_B$ operation to both sides, we have
\begin{eqnarray}
&&\rho^{\prime I}_S(t)-\rho^{\prime I}_S(0)=-i\int^t_0
dt'\mbox{Tr}_B[H'_1(t'),\rho^{\prime I}_c(t')].
\end{eqnarray}
Other terms in Eq.(27) are
\begin{eqnarray}
&&\rho^{\prime I}_c(t)=\rho^{\prime I}_c(0)-i\int^t_0
dt'[H'_1(t'),\rho^{\prime I}_S(t')\rho_B].
\end{eqnarray}
Substituting this equation into Eq.(28) we get the master equation
for $\rho^{\prime I}_{S}(t)$\cite{scu}
\begin{eqnarray}  \label{mast-eq}
&&\frac{d}{dt}\rho^{\prime I}_{S}(t)
=-i\mbox{Tr$_B$}[H'_1(t),\rho^{\prime I}_c(0)]
-\int^t_0\mbox{Tr$_B$}[H'_1(t),[H'_1(t'),\rho^{\prime
I}_S(t')\rho_B]]dt'.
\end{eqnarray}
where all higher order (than $g^2_l$) terms are neglected.


At $t=0$, the usual initial density operator is
$\rho_{SB}(0)=\left(
\begin{array}{cc}
1 & 0\\
0 & 0
\end{array}
\right) \rho_B$. Then one can get the initial condition for the
calculations: $\rho^{\prime
I}_{SB}(0)=\rho'_{SB}(0)=e^S\rho_{SB}(0)e^{-S}$ leads to
\begin{eqnarray}
&&\rho^{\prime I}_S(0)=\mbox{Tr}_B\rho'_{SB}(0)=\left(
\begin{array}{cc}
1 & 0\\
0 & 0
\end{array}
\right), \mbox{~~~}\rho^{\prime I}_c(0)=\left[S, \left(
\begin{array}{cc}
1 & 0\\
0 & 0
\end{array}
\right)\rho_B\right],
\end{eqnarray}
where we stop at the first order of $g_l$ (Eq.(\ref{S-form}) for
$S$).

The details of perturbation calculation are listed in Appendix. Note
that temperature dependence ($\tanh (\frac{\omega_l'}{2T})$) appears
explicitly in the derivation. We obtain that the diagonal elements
in the reduced density matrix are independent of temperature,
nondiagonal elements are dependent on temperature. The solution of
the reduced density operator $\rho^{\prime}_S(t)=\left(
\begin{array}{cc}
\rho^{\prime}_{11} & \rho^{\prime}_{12}\\
\rho^{\prime}_{21} & \rho^{\prime}_{22}
\end{array}
\right)$ is
\begin{eqnarray} \label{r11-r22}
&&\rho^{\prime}_{11}(t)-\rho^{\prime}_{22}(t)=\frac{1}{4\pi
i}\int^{-\infty}_{\infty} e^{-i\omega t}d\omega
\frac{1}{\omega-\eta\Delta-\sum_l\frac{V_l^2}{\omega-\omega'_l+i0^+}}\nonumber\\
&&+\frac{1}{4\pi i}\int^{\infty}_{-\infty} e^{i\omega t}d\omega
\frac{1}{\omega-\eta\Delta-\sum_l\frac{V_l^2}{\omega-\omega'_l-i0^+}}
.
\end{eqnarray}
The real and imaginary parts of $ \sum_{l} V_l^2/(\omega
-i0^{+}-\omega' _{l})$ are denoted as $R(\omega )$ and $\gamma
(\omega)$, respectively. They are
\begin{eqnarray}
R(\omega
)&=&\sum_l\frac{\eta^2\Delta^2g^2_l}{\sqrt{(\omega_l+\eta\Delta)^2+g^2_l}}
\frac{1}{\omega\sqrt{(\omega_l+\eta\Delta)^2+g^2_l}-\omega_l(\omega_l+\eta\Delta)-g^2_l}\nonumber\\
&=&-2\alpha\frac{(\eta\Delta)^2}{\omega+\eta\Delta}
\left\{\frac{\omega_c}{\omega_c+\eta \Delta}
-\frac{\omega}{\omega+\eta \Delta}
\ln\left[\frac{|\omega|(\omega_c+\eta \Delta)}
{\eta \Delta(\omega_c-\omega)}\right]\right\},\\
\gamma (\omega
)&=&\sum_l\frac{\pi\eta^2\Delta^2g^2_l}{(\omega_l+\eta\Delta)^2+g^2_l}
~~\delta\left(\omega-\frac{\omega_l(\omega_l+\eta\Delta)+g^2_l}{\sqrt{(\omega_l+\eta\Delta)^2+g^2_l}}\right)\nonumber\\
&=&2\alpha\pi\omega\frac{(\eta\Delta)^2}{(\omega+\eta\Delta)^2}.
\end{eqnarray}

The decay rate $\gamma(\omega)$ is not dependent on temperature but
on frequency which agrees with the conclusion of Nitzan and Silbey's
paper\cite{nitzan}. Its nonmonotonic behavior exhibits a maxima at
$\omega=\eta\Delta$ which is distinguished from the rotating-wave
approximation (RWA) results($\gamma_{RWA}=2\alpha\pi\omega$). The
non-equilibrium correlation function $P(t)$ is defined as
\begin{eqnarray} \label{Pt}
&&P(t)=\langle\tau_z(t)\rangle=\mbox{Tr}_S\mbox{Tr}_B(\rho_{SB}(t)\tau_z)
=\mbox{Tr}_S\mbox{Tr}_B(\rho^{\prime}_{SB}(t)\sigma_z)\nonumber\\
&&=\mbox{Tr}_S\mbox{Tr}_B\left([\rho^{\prime}_S(t)\rho_B+e^{-iH'_0t}\rho^{\prime
I}_c(t)e^{iH'_0t}] \sigma_z\right)
=\mbox{Tr}_S\left(\rho^{\prime}_S(t) \tau_z\right)\nonumber\\
&&=\rho^{\prime}_{11}(t)-\rho^{\prime}_{22}(t)=\frac{1}{\pi}\int^{\omega_c}_{0}
d\omega\frac{\gamma(\omega)\cos(\omega
t)}{[\omega-\eta\Delta-R(\omega)]^2+\gamma^2(\omega)},
\end{eqnarray}
since $\mbox{Tr$_B$}\rho_B=1$ and $\mbox{Tr$_B$}\rho^{\prime
I}_c(t)=0$. Here $\langle\tau_z(t)\rangle$ is used to denote the
average $\mbox{Tr}_S\mbox{Tr}_B(\rho_{SB}(t)\tau_z)$, which is the
population difference. The last equality in Eq.(\ref{Pt}) comes from
Eqs.(\ref{r11-r22}) with the use of the Kramers-Kronig relation.

Note that $P(t)$ is temperature-independent, which means that the
increase of temperature does not favor nor suppress the coherent
dynamics, which is distinguished from the temperature effects in the
SBB model(the coherence loses quickly  with increasing the
temperature). Fig. 1 shows the $P(t)$ versus $\Delta t$ relations
with $\Delta/\omega_c=0.1$ for different couplings $\alpha$. From
the evolution behavior of the SSB, it is found that, with increasing
coupling, the dynamics exhibits from the damped coherent-oscillation
for the weak coupling to incoherent decay for moderate coupling.
Fig. 2 shows the $P(t)$ versus $\eta\Delta t$ relations with
$\alpha=0.1$ and different tunneling. The curves show a similar
scaling behavior for different tunneling $\Delta$.

The integration in Eq.(\ref{Pt}) can be done approximately by the
residue theorem,
\begin{eqnarray}\label{WW}
&&P(t)=\cos(\omega_0 t)\exp(-\gamma t),
\end{eqnarray}
where $\omega_0$ is the solution of equation
\begin{eqnarray} \label{w0eq}
&&\omega_0-\eta\Delta-R(\omega_0)=0,
\end{eqnarray}
and $\gamma$ is the Wigner-Weisskopf approximation (WWA) of
$\gamma(\omega)$:
\begin{eqnarray}
&&\gamma =\gamma(\omega=\eta\Delta)={1\over 2}\alpha \pi
\eta\Delta .
\end{eqnarray}
Notice that $\omega_0-\Delta=(\eta-1)\Delta+R(\omega_0)$ is the
level shift induced by the spin-bath coupling. In weak coupling
case, the integral function in Eq. (\ref{r11-r22}) of $P(t)$
possesses two complex poles  which result in damped oscillation
dynamics \cite{rmp,book}. The real part $\omega_0$ represents the
frequency of coherent tunneling. With increasing coupling,
$\omega_0$ become smaller. The solution $\omega_0$ of
Eq.(\ref{w0eq}) is real ($\omega_0\ge 0$) only when
$\alpha\le\alpha_c$, and
\begin{eqnarray}\label{alpha_c}
&&\alpha_c={\frac{1}{2}}\left\{1+\frac{\eta\Delta}{\omega_c}\right\}
\end{eqnarray}
is determined by $\omega_0=0$ in Eq.(\ref{w0eq}). It becomes the
well-known result $\alpha_c=1/2$ for the SSB in the scaling limit $
\Delta/\omega_c\ll 1$. For $\alpha >\alpha _{c}$ there is no real
solution $\omega _{0}$ and it means that $\alpha = \alpha _{c}$
determines the critical point for a coherent-incoherent transition
in contrast to the diffuse boundary region between the coherent
phase and incoherent one predicted by numerical path integral
treatment\cite{Makri}. Besides, the critical coupling is independent
of temperature, so there is no boundary shift as the temperature is
raised, which is in contrast with the known behavior of an Ohmic
bath of bosons and the NIBA results with weak-temperature
dependence\cite{rmp,Makri}. Thus, it is found that the increase of
spin-bath temperature does not favor the coherence.

In Fig. 3, a phase diagram shows the relation between temperature
and coupling. In comparison with numerical path integral results,
and the NIBA results of both the SSB and SBB models,  our analytical
results show the crossover from the coherent oscillations to
incoherent decays is independent of temperature. The dashed line
shows the coherent-incoherent boundary predicted by the NIBA with an
effective harmonic bath of temperature-dependent spectral density
for $\omega_c=20\Delta$ (All data are taken from Ref.\cite{Makri}).
Thus, the NIBA solution for the spin-bath indicates that the
coherent-incoherent boundary shifts to stronger coupling as the
temperature is raised, which might be questionable since NIBA is not
reliable for small values of cutoff frequency and low temperature
regime\cite{rmp,book,Makri}. From Fig. 3, one can see that at lower
temperature, the critical coupling for the SSB obtained by the NIBA
is about $0.75$ which is larger than $\alpha_c=0.5$ for the SBB
model for $\Delta/\omega_c \ll 1$ at $T=0$. However, at zero
temperature, from the form of bath correlation functions of the two
models by the NIBA shown in the following \cite{Makri,Dekker}, the
SSB and SBB model yield identical results which is the same as the
conclusion obtained by the resolvent operator approach that there is
the same underlying physics for both two models at zero temperature
\cite{Shao}. Additionally, the simulations by numerical path
integral is also shown for comparison in the shaded area. The
parameter space is obtained by an experiential method that $P(t)$
has a small negative lobe that does not fall below $-0.01$ which is
stated in Ref.[\cite{Makri}]. Consequently, at moderate or higher
temperature, the width of this area is considerably broad whose
parameters is corresponding to the criterion \cite{Makri}. In
contrast, our approach formulates $P(t)$ for the SSB model obviously
without temperature dependence. In the scaling limit, one can
readily get $\alpha_c=0.5$ which is consistent with the exact
results\cite{book, Shao}.  As a consequence, our result turns out
that the coherence of the TLS does not benefit from the increase of
temperature.

Since Eq.(\ref{Pt}) is temperature independent, these conclusions
hold also true for the finite temperature. This is totally different
from those of SBB, as the coherent oscillation of SBB disappears
quickly with increasing temperature. On the other hand, NIBA results
for boson-bath indicate that the transition temperature drops very
quickly as the dissipation increases which is shown in Fig. 3. The
$P(t)$ of the SBB model is also calculated by our
approach\cite{zheng,lu},
\begin{eqnarray}\label{pt-sb}
&&P(t)=\rho^{\prime}_{11}(t)-\rho^{\prime}_{22}(t)=\frac{1}{\pi}\int^{\omega_c}_{0}
d\omega\frac{\gamma_B(\omega)\cos(\omega
t)}{[\omega-\eta\Delta-R_B(\omega)]^2+\gamma_B^2(\omega)},
\end{eqnarray}
where $R_B(\omega)$ and $\gamma_B(\omega)$ are the real and
imaginary part of $\sum_l V_l^2
\coth(\frac{\omega}{2T})/(\omega-\omega_k-i0^{+})$, respectively
(For boson bath, $V_l=\eta_B \Delta g_l/(\eta_B \Delta +
\omega_l)$). They are given by
\begin{eqnarray}
R_B(\omega )&=&-2\alpha \frac{(\eta_B\Delta )^{2}}{\omega +\eta_B
\Delta } \left\{ \frac{\omega _{c}}{\omega _{c}+\eta_B \Delta
}-\frac{\omega }{ \omega +\eta_B \Delta }\ln \left[ \frac{|\omega
|(\omega _{c}+\eta_B \Delta )}{\eta_B \Delta |\omega _{c}-\omega
|}\right] \right\}\nonumber\\
&&+\sum_{l}\frac{ V_{l}^2}{\omega
-\omega _{l}}\frac{2}{\exp(\omega_k/T)-1},\\
\gamma_B (\omega )&=&2\alpha \pi \omega \frac{(\eta_B \Delta
)^{2}}{(\omega +\eta_B \Delta )^{2}}\coth (\frac{\omega }{2T}),
\mbox{~~for~~}0\le\omega\le\omega_c.
\end{eqnarray}
In order to show our method explicit, we compare the result of our
approach to the results on the spin-boson model at $T=0$ from
numerical renormalization group theory\cite{anders}, which is shown
in Fig. 4. It is found that from weak coupling to moderate coupling,
our result is in good agreement with those of the numerical method.
However, there appears a difference near the coherent-incoherent
transition.

At $T=0$, the population difference of the SSB model is the same as
that of SBB model. It confirms that they exhibit the same dynamics
at zero temperature which coincide with the findings obtained by
Shao and H\"{a}nggi \cite{Shao}. The decay rate $\gamma_B(\omega)$
is temperature dependent which is consistent with the known
results\cite{book,nitzan}. By comparing $\gamma(\omega)$ in
Eq.(\ref{Pt})(without temperature factor) at finite temperature with
$\gamma_B(\omega)$ in Eq.(\ref{pt-sb})(with $\coth (\frac{\omega
}{2T})$), we find that the difference of dissipative roles between
the two kinds of baths ascribes to the two distinct level structures
for each bath degree of freedom and available states distribution.

The difference between the two models can be traced back to the
restriction of the thermal induced excitation possibilities of any
bath degrees of freedom. For spin-bath, there is only a single level
of thermal excitation in each individual two-level system of bath,
while for boson-bath, there are infinite levels in each individual
oscillating mode. In the boson-bath model, thermal excitation of
many levels of a single bath degree of freedom is one of important
mechanism for decoherence, whose excitation number can be
represented by $2 n_k+1 = \coth(\omega_k/2T)$, while the spin-bath
model lacks in this structure. Thus, our approach formulates the
decoherence measure $P(t)$ for the SSB model obviously without
temperature dependence.

$\langle\tau_x(t)\rangle$ can be calculated in a similar way as
Eq.(\ref{Pt}),
\begin{eqnarray*}
&&<\tau_x(t)>=\mbox{Tr}_S\mbox{Tr}_B(\rho^{\prime}_{SB}(t)e^{S}\tau_xe^{-S})\nonumber\\
&&=\mbox{Tr}_S\mbox{Tr}_B\left([\rho^{\prime}_S(t)\rho_B+e^{-iH'_0t}\rho^{\prime
I}_c(t)e^{iH'_0t}] [\tau_x\cosh(X)+i\tau_y\sinh(X)]\right)\nonumber\\
&&=\mbox{Tr}_S\mbox{Tr}_B\left(\rho^{\prime}_S(t)\rho_B\tau_x\cosh(X)+i\rho^{\prime
I}_c(t)e^{iH'_0t}\tau_ye^{-iH'_0t}\sinh(X(t))\right),
\end{eqnarray*}
where $X=\sum_lv_l(\sigma_{-}^{l}-\sigma_{+}^{l})$ and
$X(t)=\sum_lv_l(\sigma_{-}^{l}e^{-i\omega_lt}-\sigma_{+}^{l}e^{i\omega_lt})$.
The trace operation related to the cross term $\rho^{\prime I}_c(t)$
can be done with Eq.(A16) and the result is
\begin{eqnarray}\label{tau-x}
&&\langle\tau_x(t)\rangle =\eta\tanh(\frac{\eta\Delta}{2T})
\left\{1-\exp(-2\gamma t)\right\}
-\frac{1}{\Delta}\sum_lv_lV_l\tanh(\frac{\omega_l}{2T})\sin(\omega_lt)\sin(\eta\Delta t)\nonumber\\
&&+\frac{1}{\Delta}\sum_lV^2_l[\tanh(\frac{\omega_l}{2T})-\tanh(\frac{\eta\Delta}{2T})]
\frac{1-\cos[(\omega_l-\eta\Delta)t]}{\omega_l-\eta\Delta}\nonumber\\
&&+\frac{1}{\Delta}\sum_lV^2_l \tanh(\frac{\eta\Delta}{2T})
\frac{(\omega_l-\eta\Delta)[\exp(-2\gamma
t)-\cos[(\omega_l-\eta\Delta)t]]+2\gamma\sin[(\omega_l-\eta\Delta)t]}{(\omega_l-\eta\Delta)^2+4\gamma^2}
\end{eqnarray}
where we have taken into account the terms up to the second order of
$g_l$. Here, the temperature plays some role, and Eq. (\ref{tau-x})
leads to correct long time limit $\eta
\tanh(\frac{\eta\Delta}{2T})$. One can easily check that the initial
conditions $P(0)=1$, and $\langle\tau_x(t=0)\rangle=0$ are well
satisfied. Besides,
\begin{eqnarray}
&&\langle\tau_x(t\to\infty)\rangle=\eta\tanh(\frac{\eta\Delta}{2T}),
\mbox{~} P(t\to\infty)=0,
\end{eqnarray}
which are the correct results for thermodynamic equilibrium state.

In our work, the dynamic behavior of the central spin is mainly
determined by the real and imaginary parts of the self energy $
\sum_{l} V_l^2/(\omega -i0^{+}-\omega' _{l})$, $R(\omega )$ and
$\gamma (\omega)$. Eqs.(33) and (34) show that both $R(\omega )$ and
$\gamma (\omega)$ are temperature independent and this leads to our
conclusion that finite temperature does not favor the coherence.
However, the NIBA may lead to the population difference
$P_{NIBA}(t)$ (Refs.\cite{Makri} and \cite{Dekker}) with the Laplace
transform $P(s)$
\begin{eqnarray*}
&&P(s)=[s+f(s)]^{-1},
\end{eqnarray*}
where
\begin{eqnarray*}
&&f(s)=\Delta^2\int_0^{\infty}\cos(Q_1(t))\exp(-st-Q_2(t))
dt,\\
&&Q_1 = \int_0^{\infty} \sin(\omega t) \tanh(\frac{\omega}{2T})
\frac{J(\omega)}{\omega^2}
d\omega ,\\
&&Q_2 = \int_0^{\infty} [1-\cos(\omega t)]
\frac{J(\omega)}{\omega^2} d\omega.
\end{eqnarray*}
Nevertheless for bosonic bath, the two functions become
\begin{eqnarray*}
&&Q_1 = \int_0^{\infty} \sin(\omega t) \frac{J(\omega)}{\omega^2}
d\omega ,\\
&&Q_2 = \int_0^{\infty} [1-\cos(\omega t)]\coth(\frac{\omega}{2T})
\frac{J(\omega)}{\omega^2} d\omega.
\end{eqnarray*}
Then, Ref.28 claimed that for the spin bath one may introduce an
effective spectrum $J_{eff}(\omega)=J(\omega)\tanh(\omega/(2T))$ and
the coherent-incoherent boundary shifts to stronger coupling (larger
$\alpha$) as the temperature is raised, which is contrary to the
effect of bosonic bath (with increasing temperature, $\alpha_c$
decreases quickly).

In Ref.\cite{Shao}, a polaronic transform is used and, then, the
resolvent operator approach is applied to the transformed
Hamiltonian. It is founded that for the SSB model $P(t)$ is
effectively temperature independent at low and high temperature,
which is the same as ours. For finite temperature, with the similar
formulation as those of NIBA Ref.\cite{Shao} concludes that
temperature plays, though weakly, a positive role in maintaining
coherent dynamics, which is different from our result of temperature
independence.

Generally speaking, the polaronic transformation (Refs.\cite{Shao}
and \cite{Dekker}) and the NIBA (Ref. \cite{Makri}) lead to the
second order perturbation with perturbation parameter $\Delta$,
since $f(s)$ is explicitly proportional to $\Delta^2$ and $Q_1$ and
$Q_2$ are $\Delta$ independent. However, based on the transformed
Hamiltonian our approach is the second order perturbation with
renormalized coupling parameter $V_l$ which has included the effects
of the renormalized tunneling arising from the coupling to the bath.
For weak coupling, $V_l \simeq g_l $. In calculations we take into
account all the second order terms of $g^2_l$ but neglect higher
order terms included in $H'_2$, which are the multi-flip-flop
processes of bath spins of different modes, that is, two or more
bath spins flip of different modes at the same time. Thus, we
provide some general checks based on some known results. The
treatment can be justified by the shiba relation and sum rule.

\section{Correlation function and Shiba relation}
Since $\exp(S)\tau_z\exp(-S)=\tau_z$, the retarded Green's function
can be written as
\begin{eqnarray}
&&G(t)=-i\theta (t)Z^{-1}\mbox{Tr}\left\{ \exp (-\beta H')\{\exp
(iH't)\tau_{z}\exp (-iH't),\tau_{z}\}\right\} ,
\end{eqnarray}
where $H'\approx H'_0+H'_1$ is the transformed Hamiltonian. The
usual notation for the Fourier transform of $G(t)$ is $ G(\omega
)=\langle \langle \tau_{z};\tau_{z}\rangle \rangle $, where $\langle
\langle A;B\rangle \rangle$ denotes the retarded Green's function
for operators $A$ and $B$ which satisfies the following equation of
motion,
\begin{eqnarray}
&&\omega \left\langle \left\langle A;B\right\rangle \right\rangle
=\left\langle \{A,B\}\right\rangle ^{\prime }+\left\langle
\left\langle
[A,H^{\prime }];B\right\rangle \right\rangle ,  \nonumber \\
&&\left\langle \{A,B\}\right\rangle ^{\prime
}=Z^{-1}\mbox{Tr}\left\{ \exp (-\beta H^{\prime })\{A,B\}\right\} .
\nonumber
\end{eqnarray}
So the solution for $G(\omega )$ is
\begin{eqnarray}
&&G(\omega )=\frac{1}{\omega -\eta\Delta -\sum_{l}V_{l}^{2}/(\omega
-\omega' _{l})} +\frac{1}{\omega +\eta\Delta
-\sum_{l}V_{l}^{2}/(\omega +\omega' _{l})}.
\end{eqnarray}
Thus, the symmetrized correlation function
\begin{eqnarray}
&&C(t)=\frac{1}{2}\mbox{Tr}\left\{\exp(-\beta H)[\tau_{z}(t)\tau_z
+\tau_z\tau_z(t)]\right\} /Z  \nonumber \\
&&=-\frac{1}{2\pi}\int^{\infty}_{-\infty}d\omega \mbox{Im}
G(\omega)\exp(-i\omega t)  \nonumber \\
&&=\frac{1}{\pi}\int^{\infty}_0 d\omega \frac{\gamma(\omega)}
{[\omega -\eta\Delta-R(\omega)]^2+\gamma^2(\omega)}\cos(\omega t).
\end{eqnarray}
The susceptibility at zero temperature is related to the Green's
function by $\chi^{\prime\prime}(\omega )=-\mbox{Im}G(\omega
)\mbox{Sgn}(\omega)$ as follows,
\begin{eqnarray}
&&\chi^{\prime\prime}(\omega )=\frac{\gamma (\omega )\theta (\omega
)}{[\omega -\Delta \eta-R(\omega )]^{2}+\gamma ^{2}(\omega )}
-\frac{\gamma (-\omega )\theta (-\omega )}{[\omega +\Delta \eta
+R(-\omega )]^{2}+\gamma ^{2}(-\omega )}.
\end{eqnarray}
The static susceptibility $\chi_0$ can be extracted with a
Kramer's-Kronig relation and a function-dissipation theorem
\begin{eqnarray}
&&\chi_{0}=\frac{2}{\pi}\int_0^{\infty}d\omega
\frac{\chi^{\prime\prime}(\omega)}{\omega}.
\end{eqnarray}
One can check the Shiba-relation\cite{wei,vol,cos,ks}:
\begin{equation}
\lim_{\omega \rightarrow 0}\frac{ C(\omega )}{J(\omega) }=
(\frac{\chi_{0}}{2})^{2}.
\end{equation}
Note that our normalization condition $\int_0^{\infty}d\omega
C(\omega)=1$, i.e. the sum rule is another check for our approach.
We provide results for the Shiba-relation and sum rule in the Ohmic
case for various values $\alpha$ and $\Delta$ in Table 1. It turns
out that the Shiba-relation is exact satisfied in numerical
precision in the coherent regime. Outside the regime, the agreement
is still good but no longer exact. Approximations schemes like NIBA
or numerical methods based on Monte Carlo cannot be used to verify
the Shiba relation since they fail to predict the correct long-time
behavior\cite{stauber}.

\section{Summary and discussion}
The physics of the SSB model is studied by means of the perturbation
approach based on a unitary transformation. Analytical results of
the quantum dynamics, described by the reduced density operator
$\rho(t)$, is obtained for both the scaling limit $\Delta /\omega
_{c}\ll 1$ and the general finite $\Delta /\omega _{c}$ case. $P(t)$
is temperature independent while $<\tau_x(t)>$ is temperature
dependent. Moreover, the decay rate is temperature independent which
is in good agreement with the conclusion of Nitzan and Silbey's
paper\cite{nitzan}. It is found that the transition from coherent to
incoherent dynamics happens at
$\alpha_c={\frac{1}{2}}[1+\eta\Delta/\omega_c]$, which is
temperature independent. Our results have answered the two problems
mentioned in the introduction. Even though the SSB model has the
same dynamics as the SBB model at $T=0$ in the coherent regime, they
displays distinctive dynamics at $T>0$. Furthermore, in the
boson-bath, the population difference decreases fast with increasing
temperature, while it is independent of temperature in the
spin-bath. Besides, the dynamical properties obtained by our
approach can both well satisfy initial conditions and reasonably
obtain the thermodynamical limits. The conclusion that the coherent
oscillation (the population difference) does not depend the
temperature for the SSB model is not a bad news to the study of
quantum information processing in the low temperature regime by
nanomagnets and nuclear spins.

If $P(t)$ is calculated by the polaron transformation, the second
perturbation theory in the tunneling matrix element is applied.
Evaluations about $P(t)$ reproduces the same expression of the NIBA
which is seen in Ref.[\cite{Dekker}](H. Dekker, Phys. Rev. A35,
1436(1987)). Note that the NIBA self-energy function is the second
order $\Delta$ because there only exists a prefactor $\Delta^2$ in
Eq.(4), both $Q_1$ and $Q_2$ are independent of $\Delta$.  However,
our approach is the combined second order perturbation with
renormalized coupling $V_l$ which has included the effects of
tunneling $\Delta$ and coupling to the bath $\alpha$. On one hand,
the contribution of higher order $\Delta$ has been taken into
account due to the form of $\xi_k$ with renormalized tunneling.
Physically, the self energy function $ \sum_{l} V_l^2/(\omega
-i0^{+}-\omega' _{l})$ with the renormalizied coupling $V_l$ has
been included the higher order of $\Delta^2$, whose real part and
imaginary one exhibit it explicitly. On the other hand, the coupling
to bath leads to diagonal transitions and off-diagonal transitions.
All diagonal transitions have been accounted by $\eta$ in $H'_0$,
while non-diagonal transitions have been considered to the first
order $V_l$ which has been included into $H'_1$ and give effects of
order $\alpha$. Obviously, if we make $\xi_k=1$ for any $k$ , the
same conclusion as the NIBA can be drawn.  Since our transformation
is different from the usual polaron transformation, we come to
different conclusions.

In our treatment two approximations are applied. One is the omission
of the perturbation term $H'_2$ corresponding to multi-phonon
non-diagonal transitions such as $b^{\dag}_k b^{\dag}_{k'}$ in the
spin-boson bath model or multi-bath-spin flips on different modes at
the same time such as $\sigma^{+}_l\sigma^{+}_{l'}$ in the
spin-spin-bath model due to its contribution to physical quantities
$O(g_k^4)$ or higher orders. This approximation can be justified by
the Shiba relation and sum rule (or the initial condition
$P(t=0)=1$). The other approximation is usual Born approximation for
deriving the master equation. The generalized Shiba relation and sum
rule provide a good test of our method, not just at low frequency,
but at all energy scales and for values of $\alpha$ beyond weak
coupling and a finite $\Delta$.

The quantum dynamics of both the SSB model and SBB model are studied
analytically by the perturbation method based on a unitary
transformation. Our approach can be justified by our numerical
results: (1)The long-time limits of $P(t)$ and
$\langle\tau_x(t)\rangle$ are correct, namely, the expectation value
$\langle\tau_z(\infty)\rangle$ vanishes as expected and
$\langle\tau_x(\infty)\rangle$ goes to the value of thermodynamical
equilibrium. Besides, the initial conditions are correct. (2)The
coherent-incoherent transition point $\alpha_c=0.5$ is the same as
the known result of previous authors. (3)The shiba's relations of
both SBB and SSB models with Ohmic spectrum have been checked well
within a wide parameter range (to see Table I and our previous work
Ref.\cite{lu}).

Here are a few words about the key ingredient of the approach. The
key point of our treatment is the unitary transformation with
generator Eq.(7), where a parameter $\xi_k$ is introduced. After the
transformation a perturbation expansion has been performed. If
$\xi_k=0$ for any $k$, that is, without the transformation, the
perturbation expansion would be similar to the standard
weak-coupling expansion (Bloch-Redfield theory). In addition, if
$\xi_k=1$ for any $k$, then our transformation is the usual
polaronic transformation and the perturbation expansion is for the
small parameter $\Delta$ which is equivalent to the NIBA (H.Dekker,
PRA35, 1436(1987)). Our choice for $0<\xi_k<1$ (Eq.(14)) is between
them and thus is an improvement on the analytical methods.

The purpose of our unitary transformation is to find a better way to
divide the transformed Hamiltonian into unperturbed part
$H^{\prime}_0$, which can be treated exactly, and perturbation ones
$H^{\prime}_1+H^{\prime}_2$, which may be treated by perturbation
theory. If one treats the coupling term in the original Hamiltonian
$H$ as the perturbation, the dimensionless expanding parameter is
$g_{l}^{2}/\omega _{l}^{2}$. For Ohmic bath $s=1$ it is
$2\alpha/\omega $ which is logarithmic divergent in the infrared
limit. By choosing the form of $\eta $ (Eq.(\ref{eta-s})) and
introducing the function $\xi_k$ in the unitary transformation it is
possible to treat $H_{1}^{\prime }$ and $H_{2}^{\prime }$ as
perturbation because of the following reason. On account of the form
of $\eta$ in Eq.(\ref{eta-s}) $H_{2}^{\prime}$ can be treated as
perturbation because its contribution is zero at second order of
$g_l$. The effect of the coupling term in $H^{\prime}$ ($
H^{\prime}_1$) can be safely treated by perturbation theory because
the infrared divergence in the original perturbation treatment for
$H$ is eliminated by making choice of the function form $\xi_l$. The
expanding parameter ($s=1$) is $g_{l}^{2}\xi _{l}^{2}/\omega
_{l}^{2}\sim 2\alpha\omega /(\omega +\eta \Delta )^{2}$, which is
finite in the infrared limit. This approach works well for the
low-temperature coherent region and the tunneling $
0<\Delta<\omega_c $. It is quite tractable and physically clear, it
produces nearly all results which agree with exact ones obtained by
various complicated methods in the SSB and SBB model. Thus it may be
easily extended to more complicated coupling systems.

\vskip 0.5cm

{\noindent {\large {\bf Acknowledgement}}}

\bigskip  We would like to thank Prof. Frithjof B. Anders for providing the data of numerical renormalization group. We would
like to acknowledge the support from the National Science Foundation
of China under grant Nos. 90503007 and 10547126.

\section*{Appendix}

\setcounter{equation}{0}
\renewcommand{\theequation}{A\arabic{equation}}

In this Appendix we list the details of solving the master equation
(\ref{mast-eq}). The first term at right side of Eq.(\ref{mast-eq})
is
\begin{eqnarray}
&&-i\mbox{Tr$_B$}[H'_1(t),\rho^{\prime I}_c(0)]=
-i\mbox{Tr$_B$}[H'_1(t),[S,\left(
\begin{array}{cc}
1 & 0\\
0 & 0
\end{array}
\right)\rho_B]]\nonumber\\
&&=-i\mbox{Tr$_B$}[H'_1(t),{1\over 2}\sum_lv_l\left(
\begin{array}{cc}
1 & 0\\
0 & 0
\end{array}
\right)\left\{i\sigma_y^{l}\rho_B-\rho_Bi\sigma_y^{l}\right\}]\nonumber\\
&&=\frac{1}{2\eta\Delta}\sum_lV^2_l\tanh(\frac{\omega'_l}{2T})\tau_x\sin[(\omega'_l-\eta\Delta)t].
\end{eqnarray}

The integration in Eq.(\ref{mast-eq}) can be done as follows,
\begin{eqnarray}
&&-\int^t_0\mbox{Tr$_B$}[H'_1(t),[H'_1(t'),\rho^{\prime
I}_S(t')\rho_B]]dt'\nonumber\\
&&=-{1\over 2}\sum_lV^2_l\int^t_0dt'
\left\{\left[[1-\tanh(\frac{\beta\omega'_l}{2})][\tau_{-}\tau_+\rho^{\prime
I}_S(t')-\tau_{+}\rho^{\prime
I}_S(t')\tau_{-}]\right.\right.\nonumber\\
&&-\left.[1+\tanh(\frac{\beta\omega'_l}{2})][\tau_{-}\rho^{\prime
I}_S(t')\tau_{+}-\rho^{\prime
I}_S(t')\tau_{+}\tau_{-}]\right]\exp[i(\omega'_l-\eta\Delta)(t-t')]\nonumber\\
&&+\left[[1+\tanh(\frac{\beta\omega'_l}{2})][\tau_{+}\tau_-\rho^{\prime
I}_S(t')-\tau_{-}\rho^{\prime
I}_S(t')\tau_{+}]\right.\nonumber\\
&&-\left.\left.[1-\tanh(\frac{\beta\omega'_l}{2})][\tau_{+}\rho^{\prime
I}_S(t')\tau_{-}-\rho^{\prime
I}_S(t')\tau_{-}\tau_{+}]\right]\exp[-i(\omega'_l-\eta\Delta)(t-t')]\right\}.
\end{eqnarray}
Thus, Eq.(\ref{mast-eq}) can be solved by the Laplace
transformation,
\begin{eqnarray}
&&p\rho^{\prime I}_S(p)-\rho^{\prime I}_S(0)=
\frac{1}{2\eta\Delta}\sum_lV^2_l\tau_x
\frac{\omega'_l-\eta\Delta}{p^2+(\omega'_l-\eta\Delta)^2}\nonumber\\
&&-\sum_lV^2_l\left\{\left[[1-\tanh(\frac{\beta\omega'_l}{2})][\tau_{-}\tau_+\rho^{\prime
I}_S(p)-\tau_{+}\rho^{\prime
I}_S(p)\tau_{-}]\right.\right.\nonumber\\
&&-\left.[1+\tanh(\frac{\beta\omega'_l}{2})][\tau_{-}\rho^{\prime
I}_S(p)\tau_{+}-\rho^{\prime
I}_S(p)\tau_{+}\tau_{-}]\right]\frac{1}{p-i(\omega'_l-\eta\Delta)}\nonumber\\
&&+\left[[1+\tanh(\frac{\beta\omega'_l}{2})][\tau_{+}\tau_-\rho^{\prime
I}_S(p)-\tau_{-}\rho^{\prime
I}_S(p)\tau_{+}]\right.\nonumber\\
&&-\left.\left.[1-\tanh(\frac{\beta\omega'_l}{2})][\tau_{+}\rho^{\prime
I}_S(p)\tau_{-}-\rho^{\prime
I}_S(p)\tau_{-}\tau_{+}]\right]\frac{1}{p+i(\omega'_l-\eta\Delta)}\right\}.
\end{eqnarray}
If we denote
\[
\rho^{\prime I}_S(p)=\left(
\begin{array}{cc}
\rho^{\prime I}_{11} & \rho^{\prime I}_{12}\\
\rho^{\prime I}_{21} & \rho^{\prime I}_{22}
\end{array}
\right),
\]
the solution of Eq.(A3) is
\begin{eqnarray}
&&\rho^{\prime I}_{11}+\rho^{\prime I}_{22}=\frac{1}{p},\\
&&\rho^{\prime I}_{11}-\rho^{\prime
I}_{22}=\frac{1/2}{p+\sum_l\frac{V_l^2}{p+i(\omega'_l-\eta\Delta)}}+\frac{1/2}
{p+\sum_l\frac{V_l^2}{p-i(\omega'_l-\eta\Delta)}},\\
&&\rho^{\prime I}_{12}-\rho^{\prime
I}_{21}=\frac{1/2}{p+\sum_l\frac{V_l^2}{p+i(\omega'_l-\eta\Delta)}}-\frac{1/2}
{p+\sum_l\frac{V_l^2}{p-i(\omega'_l-\eta\Delta)}},\\
&&\rho^{\prime I}_{12}+\rho^{\prime I}_{21}=\frac{\sum_l
V^2_l\tanh(\frac{\beta\omega'_l}{2})\frac{\omega'_l+\eta\Delta}{\eta\Delta}\frac{1}
{p^2+(\omega'_l-\eta\Delta)^2}}{p\left(1+2\sum_l\frac{V_l^2}
{p^2+(\omega'_l-\eta\Delta)^2}\right)}.
\end{eqnarray}
Using the relation between $\rm {Schr\ddot{o}edinger}$ and
interaction representation (23) and making the Laplace
inverse-transformation, we can get
\begin{eqnarray}
&&\rho^{\prime}_{11}(t)+\rho^{\prime}_{22}(t)=\rho^{\prime I}_{11}(t)+\rho^{\prime I}_{22}(t)=1,\\
&&\rho^{\prime}_{11}(t)-\rho^{\prime}_{22}(t)=\cos(\eta\Delta
t)(\rho^{\prime I}_{11}(t)-\rho^{\prime
I}_{22}(t))-i\sin(\eta\Delta t)(\rho^{\prime
I}_{12}(t)-\rho^{\prime I}_{21}(t))\nonumber\\
&&=\frac{1}{4\pi i}\int e^{pt}dp\left
\{\frac{1}{p+i\eta\Delta+\sum_l\frac{V_l^2}{p+i\omega'_l}}+\frac{1}
{p-i\eta\Delta+\sum_l\frac{V_l^2}{p-i\omega'_l}}\right\},\nonumber\\
\\
&&\rho^{\prime}_{12}(t)-\rho^{\prime}_{21}(t)=\cos(\eta\Delta
t)(\rho^{\prime I}_{12}(t)-\rho^{\prime
I}_{21}(t))-i\sin(\eta\Delta t)(\rho^{\prime
I}_{11}(t)-\rho^{\prime I}_{22}(t))\nonumber\\
&&=\frac{1}{4\pi i}\int
e^{pt}dp\left\{\frac{1}{p+i\eta\Delta+\sum_l\frac{V_l^2}
{p+i\omega'_l}}-\frac{1}{p-i\eta\Delta+\sum_l\frac{V_l^2}{p-i\omega'_l}}\right\},\nonumber\\
\\
&&\rho^{\prime}_{12}(t)+\rho^{\prime}_{21}(t)=\rho^{\prime
I}_{12}(t)+\rho^{\prime I}_{21}(t)\nonumber\\
&&=\frac{1}{2\pi i}\int
e^{pt}dp\frac{\sum_lV^2_l\tanh(\frac{\beta\omega'_l}{2})\frac{\omega'_l+\eta\Delta}{\eta\Delta}\frac{1}
{p^2+(\omega'_l-\eta\Delta)^2}}{p\left(1+2\sum_l\frac{V_l^2}{p^2+(\omega'_l-\eta\Delta)^2}\right)}.
\end{eqnarray}
The integration path is on a line parallel to the imaginary axis
of complex $p$ plane from $p=0^{+}-i\infty$ to $p=0^{+}+i\infty$.
The integration in (A11) can be re-written as
\begin{eqnarray}
&&\frac{1}{2\pi i}\int e^{pt}dp \frac{\sum_l
V^2_l\tanh(\frac{\beta\omega'_l}{2})\frac{\omega'_l+\eta\Delta}{\eta\Delta}
\left[\frac{1}{p+i(\omega'_l-\eta\Delta)}
+\frac{1}{p-i(\omega'_l-\eta\Delta)}\right]}
{p\left\{p+\sum_lV_l^2\left[\frac{1}{p+i(\omega'_l-\eta\Delta)}
+\frac{1}{p-i(\omega'_l-\eta\Delta)}\right]\right\}}\nonumber\\
&&=\frac{1}{2\pi i}\int e^{pt}dp \frac{\sum_l
V^2_l\tanh(\frac{\beta\omega'_l}{2})\frac{\omega'_l+\eta\Delta}{2\eta\Delta}
\left[\frac{1}{p+i(\omega'_l-\eta\Delta)}
+\frac{1}{p-i(\omega'_l-\eta\Delta)}\right]} {\sum_l
V_l^2\left[\frac{1}{p+i(\omega'_l-\eta\Delta)}
+\frac{1}{p-i(\omega'_l-\eta\Delta)}\right]}\nonumber\\
&&\times\left\{\frac{1}{p}-\frac{1}{p+\sum_l
V_l^2\left[\frac{1}{p+i(\omega'_l-\eta\Delta)}
+\frac{1}{p-i(\omega'_l-\eta\Delta)}\right]}\right\}.
\end{eqnarray}
The pole point of the first term in $\{...\}$ is $0$ with the
residue $\tanh(\eta\Delta/2T)$. The pole point of the second term
is a real number, which can be determined by letting $1/[p+
i(\omega'_l-\eta\Delta)]+1/[p-i(\omega'_l-\eta\Delta)]=2p/[p^2+
(\omega'_l-\eta\Delta)^2]=2\pi\delta(\omega'_l-\eta\Delta)$ in
$\sum_l$ summation. Thus,
\begin{eqnarray}
&&\rho^{\prime}_{12}(t)+\rho^{\prime}_{21}(t)=\tanh(\frac{\eta\Delta}{2T})\left(1-\exp(-2\gamma
t)\right),
\end{eqnarray}
where
\begin{eqnarray}
&&\gamma ={1\over 2}\alpha \pi \eta\Delta.
\end{eqnarray}

The integration in (A8) and (A9) is more complicated than that of
(A10), because the pole points in complex $p$ plane have non-zero
real and imaginary part. We change the integration variable in (A9)
and (A10) from $p$ to $\omega$: $p=0^{+}+i\omega=i(\omega-i0^{+})$,
and the result of (A9) is listed in Eq.(\ref{r11-r22}). In order to
perform the calculation in Eq.(29) the expression for $\rho^{\prime
I}_S(t)$ is should be given by
\begin{eqnarray}
&&\rho^{\prime I}_S(t)={1\over 2}\left\{1+(1-e^{-2\gamma
t})\tanh(\frac{\eta\Delta}{2T})\tau_x\right.\nonumber\\
&&+\left.\frac{1}{\pi}\int^{\omega_c}_{0}
\frac{\gamma(\omega)d\omega}{[\omega-\eta\Delta-R(\omega)]^2+\gamma^2(\omega)}
\left[e^{-i(\omega-\eta\Delta)t}\tau_{+}+e^{i(\omega-\eta\Delta)t}\tau_{-}\right]\right\}.
\end{eqnarray}
Then, the cross term is
\begin{eqnarray}
&&\rho^{\prime I}_c(t) =\left[S, \left(
\begin{array}{cc}
1 & 0\\
0 & 0
\end{array}
\right)\rho_B\right]\nonumber\\
\nonumber\\
&&+i \sum_{l}V_{l}\int_{0}^{t}dt'\left\{ \left[\tau_{+}\sigma^{l}
_{-},\rho^{\prime
I}_S(t')\rho_B\right]e^{-i(\omega'_l-\eta\Delta)t'}+\left[\tau_{-}\sigma^{l}
_{+},\rho^{\prime I}_S(t')\rho_B\right]
e^{i(\omega'_l-\eta\Delta)t'}\right\}.
\end{eqnarray}

{\rm \baselineskip 20pt }

{\rm \newpage }

\begin{center}
{\rm {\Large {\bf Figure Captions }} }
\end{center}

{\rm \vskip 0.5cm }

{\rm \baselineskip 20pt }

%
%
%

{\rm {\bf Fig. 1}~~~$P(t)$ as a function of $\Delta t$ for
$\Delta/\omega_c=0.1$ and different tunneling $\alpha=0.05$ (solid
line), $0.1$ (dashed line), and $0.25$ (dotted line).

{\rm \vskip 0.5cm }

{\rm {\bf Fig. 2}~~~$P(t)$ as a function of $\eta\Delta t$ for
$\alpha=0.1$ and different tunneling $\Delta/\omega_c=0.01$ (solid
line), $0.05$ (dashed line), $0.1$ (dotted line) and $0.2$
(dashed-dotted line).

{\rm \vskip 0.5cm }

{\rm {\bf Fig. 3}~~~ Phase diagram for showing the relation between
temperature and coupling for the SSB model in comparison with the
results of SSB and SBB models by the NIBA. The shaded area is those
results by numerical path integral method, which indicates that the
coherent-incoherent boundary is diffuse. The solid line is our
result obtained by our method. The NIBA results for spin bath is
shown in the dashed line and those for boson-bath in dashed dotted
line. In addition, the inset displays the coherent-incoherent
boundary with relation between $\alpha_c$ and $\Delta/\omega_c$ for
the SSB model.

{\rm {\bf Fig. 4}~~~$S_z(t)=P(t)/2$ as a function of $\omega_c t$
for $\Delta/\omega_c=0.1$ and different coupling $\alpha=0.05$
(solid line), $0.1$ (dashed line), $0.25$ (dotted line) and
$\alpha_c= 0.5121$ (double-dotted dashed line).  The data of
numerical renormalization group are also shown for comparison in red
smooth lines.

\newpage
\begin{center}
{\Large \bf Table }

\vskip 0.5cm

\baselineskip 16pt

\begin{tabular}{ccccccccc}
\hline \hline & $\frac{\Delta}{\omega_c}$ & $\alpha$& $
\frac{\chi_0}{2} $ &
$\frac{C(\omega)}{J(\omega)}|_{\omega\rightarrow 0} $ & R & $C(t=0)$
\\ \hline
&0.01 & 0.1&186.5516 &34801.53&1&1&\\
&0.01 & 0.3&1170.505 &1370082&1&1&\\
&0.05 & 0.01&20.82378 &433.6306&1&1&\\
&0.05 & 0.2&54.64956&2986.575&1&1&\\
&0.05 & 0.3&116.1330&13486.87&0.9999997&1&\\
&0.05 & 0.4&366.0538&133995.4&1&1&\\
&0.1  & 0.1&14.42314&208.0271&0.9999999&0.9999992&\\
&0.1  & 0.2&22.86603  &522.8555&1&1&\\
&0.1  & 0.3&42.4048 &1798.168&1.0000005&1&\\
&0.1  & 0.4&108.7866 &11834.51&0.9999978&1&\\
&0.1  & 0.5&1536.489  &2360800&1&1&\\
&0.2  & 0.5&130.1218&16931.70&1.000001&1.000003&\\
&0.3  & 0.5&37.01318  &1369.976&1&0.9999995&\\
\hline\hline
\end{tabular}

\end{center}

\newpage
\begin{center}
{\Large \bf Table Captions }
\end{center}

\vskip 0.5cm

\baselineskip 16pt

{\bf TABLE I}: Representative results from the numerical solution
with parameters chosen by the spectral density
$J(\omega)=2\alpha\omega\Theta(\omega_c-\omega)$ and with the
controlling precision $10^{-5}$ for iteration. $R\equiv[\lim_{\omega
\to 0}C(\omega)/J(\omega)]/(\chi_0/2)^2$. The numeric error for the
Shiba relation and sum rule is at least less than $10^{-6}$ and can
be improved by increasing the accuracy of numerical calculations.

\newpage
\begin{figure}[t]
\epsfig{file=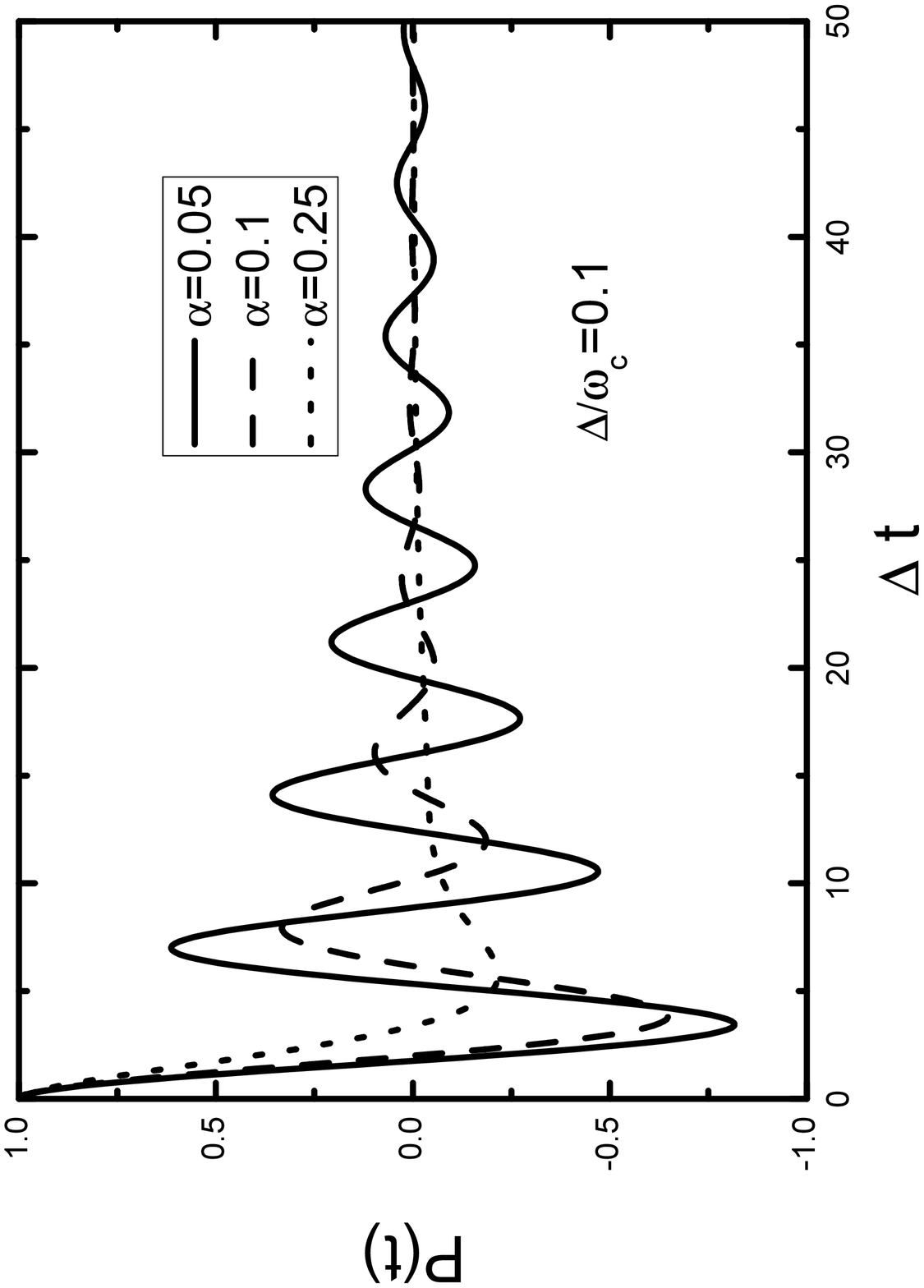,width=185mm}
\caption{\label{fig1} }
\end{figure}

\newpage
\begin{figure}[t]
\epsfig{file=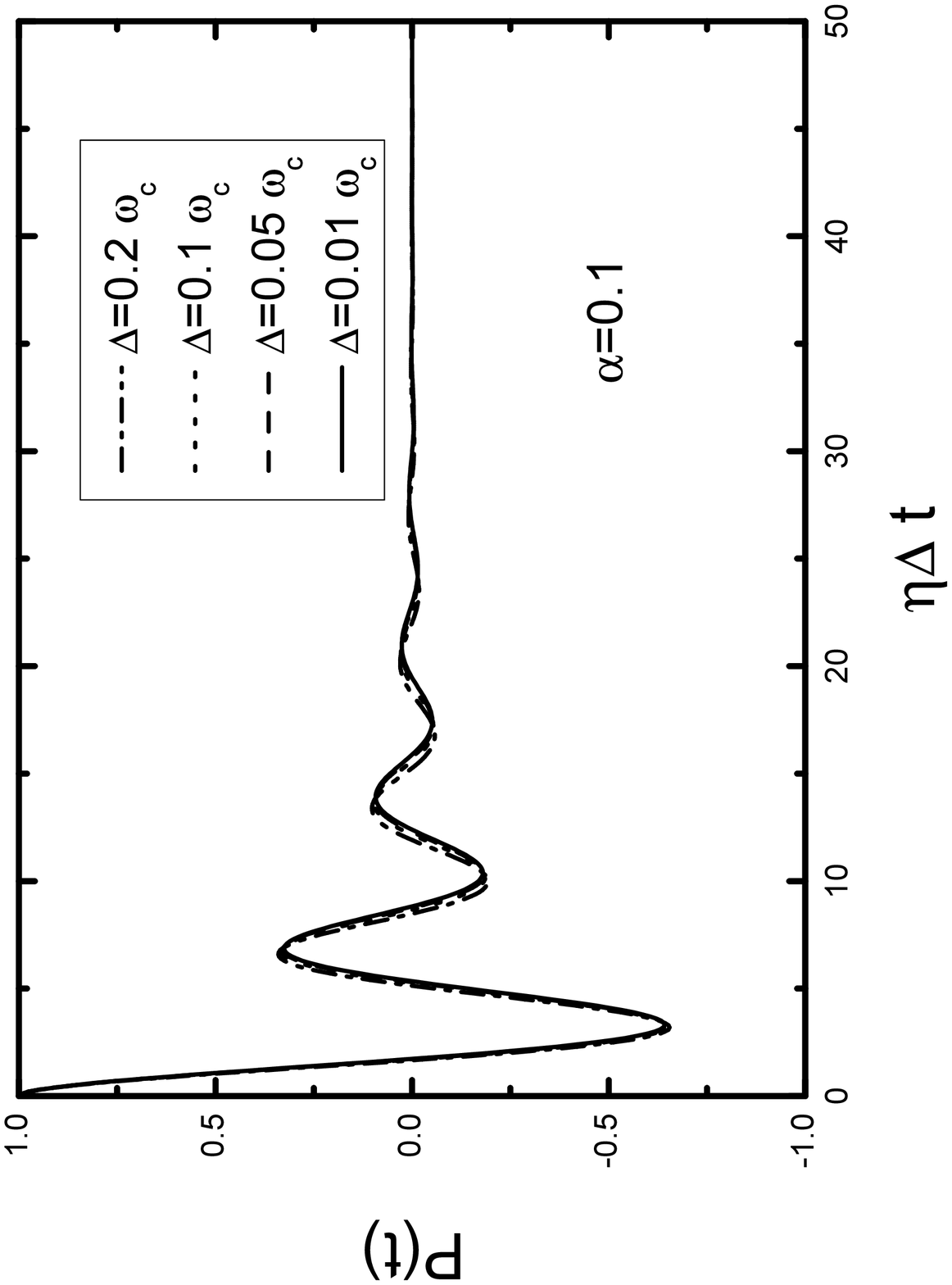,width=185mm}
\caption{\label{fig1} }
\end{figure}

\newpage
\begin{figure}[t]
\epsfig{file=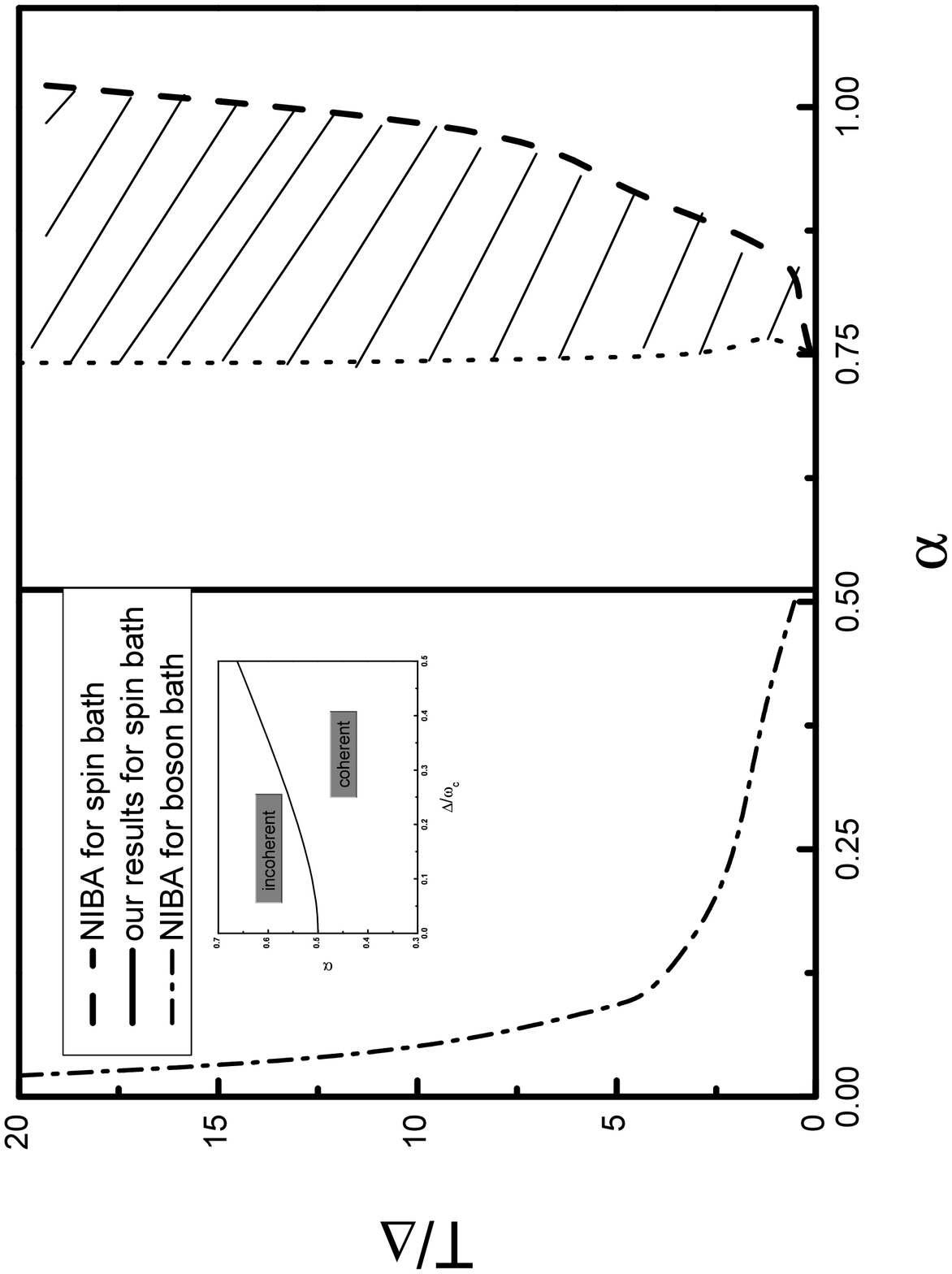,width=185mm}
\caption{\label{fig1} }
\end{figure}

\newpage
\begin{figure}[t]
\epsfig{file=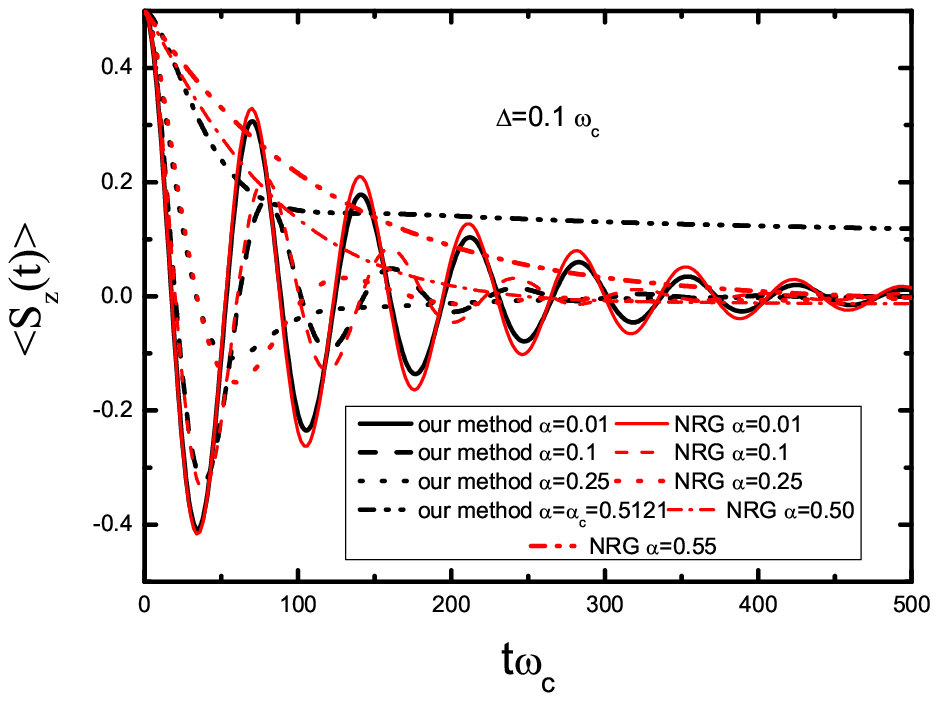,width=185mm}
\caption{\label{fig1} }
\end{figure}

\end{document}